\documentclass[aps,prb,twocolumn,superscriptaddress,nobibnotes,amsmath,amssymb,reprint]{revtex4-2}

\usepackage{graphicx}
\usepackage{xcolor}
\usepackage[normalem]{ulem}
\usepackage{natbib}

\begin{document}
\title{Local thermometry of NbSe$_2$ flake with delta-$T$ noise measurements}
\author{M.G.~Prokudina}
\affiliation{Osipyan Institute of Solid State Physics, Russian Academy of Sciences, Chernogolovka, Moscow region, 142432 Russia}
\author{A.F.~Shevchun}
\affiliation{Osipyan Institute of Solid State Physics, Russian Academy of Sciences, Chernogolovka, Moscow region, 142432 Russia}
\author{V.S.~Khrapai}
\affiliation{Osipyan Institute of Solid State Physics, Russian Academy of Sciences, Chernogolovka, Moscow region, 142432 Russia}
\author{E.S.~Tikhonov}	
\email[e-mail:]{tikhonov@issp.ac.ru}
\affiliation{Osipyan Institute of Solid State Physics, Russian Academy of Sciences, Chernogolovka, Moscow region, 142432 Russia}

\begin{abstract}
We perform transport and noise measurements for device consisting of a thin NbSe$_2$ flake laid onto the predefined gold electrodes and covered with a thin hBN~flake. In the shot noise of a~NbSe$_2$/Au tunnel junction~(TJ), we identify Andreev reflection regime by demonstrating the effective charge doubling. Further, by creating temperature gradient across the~TJ and measuring its delta-$T$ noise in the normal state, we extract electron-phonon scattering length in NbSe$_2$ and its $T$-dependence. The results of delta-$T$ noise measurements in the absence of a magnetic field when the flake is superconducting are in qualitative agreement with expectations. The introduced approach is promising for the study of nonequilibrium configurations in superconductors.
\end{abstract}
\maketitle

\section{Introduction}
Understanding of quasiparticle dynamics in superconductors is essential for improving the operation of superconductor based electronic microrefridgerators~\cite{Giazotto2006}, photon detectors~\cite{Goltsman2001,Walsh2021}, kinetic inductance elements~\cite{Grunhaupt2018} and qubits~\cite{Serniak2018}. In a static measurement, one of the convenient approaches to the study of relaxation utilizes lithographically defined tunnel junctions (TJs) which can either provide local details of the electronic states via conductance tunnel spectroscopy~\cite{Alegria2021}, or may serve as a local injector of high-energy quasiparticles~\cite{Kaplan1977}. The most advanced methods exploit cryogenic scanning microscope techniques rather than lithographic TJs~\cite{Jalabert2023}. 

In the present manuscript, we implement the approach for local thermometry of electronic states in a nonequilibrium superconductor based on the noise measurements of a weakly coupled tunnel probe which serves as an energy preserving sensor~\cite{Gramespacher1999}. Previously, similar approach was used for local thermometry of electronic states in diffusive metallic conductors using as a sensor InAs~nanowires with negligible electron-phonon scattering~\cite{Tikhonov2016}, and consequently for energy resolution of electronic states in diffusive metallic wires with sensor represented by a~TJ~\cite{Tikhonov2020}. However, up to now this approach has never been applied to superconductors. We note that this approach doesn't rely on any spectral features of superconductor or the sensor so that its applicability is not limited by high temperatures or magnetic fields~($B$) like in tunnel conductance spectroscopy experiments~\cite{Pothier1997}.

Recently, along with conventional bulk superconducting materials, transition-metal dichalcogenides~(TMDC) superconductors have also achieved much attention due to the high potential in nanoelectronics and nanophotonics applications~\cite{Qiu2021}. TMDC~layered materials offer an accessible platform with the possibility to go down to the few nanometers thickness range, assembling the devices using the now widespread stacking technique. One of the most extensively studied superconducting materials from the TMDC family is niobium diselenide. We introduce our local thermometry approach using as a testbed the thin NbSe$_2$ flake which allows us to extract the electron-phonon coupling and the temperature~($T$) dependence of electron-phonon scattering length. Our findings are relevant for the possible nonequilibrium applications of~NbSe$_2$~\cite{Shein2024}. This approach can be used for the characterization of not only the other members of TMDC superconductors down to the $2$D~limit but also for the characterization of the bulk superconductors.

\begin{figure}[!h]
\begin{center}
\includegraphics[width=\columnwidth]{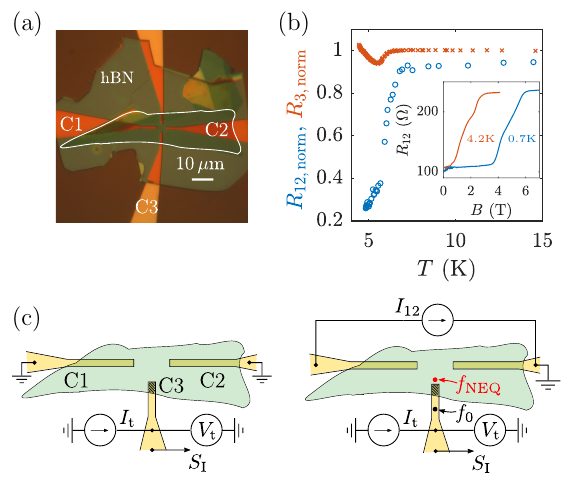}
\end{center}
\caption{(a)~Optical micrograph of the device. Exfoliated NbSe$_2$ flake (marked with white line) is put on pre-patterned gold contacts and is then covered with hBN flake. Working contacts are marked with C$1$, C$2$ and C$3$. (b)~Temperature dependence of the normalized resistances $R_{12}$ (blue circles) and $R_{3}$ (red crosses). $R_{12}$ was measured with a current of $0.45\,\mu\text{A}$ which is smaller than $0.2\Delta_0/eR_{12,N}$. $R_3$ was measured with a current of $90\,\text{nA}$ which is smaller than $0.5\Delta_0/eR_{3,N}$. Here $\Delta_0$ is the zero-$T$ superconducting gap of NbSe$_2$, $R_{12,N}$ and $R_{3,N}$ are the corresponding normal state resistances. The inset demonstrates magnetic field dependence of~$R_{12}$ at $T=4.2\,\text{K}$ (red) and $0.7\,\text{K}$ (blue). (c)~Measurement scheme for characterization of the tunnel junction realized at C$3$/NbSe$_2$ interface (left) and for conductance/noise spectroscopy (right).}
\label{fig1}
\end{figure}

\section{Device fabrication and its $T$- and $B$-behavior}
Thin NbSe$_2$ flake is first mechanically exfoliated from commercial bulk crystal (HQgraphene) onto polydimethylsiloxane stamp and is then transferred to the prepatterned $30\,\text{nm}$ thick gold contacts on an oxidized silicon chip. The same procedure is then repeated for the hBN flake which is deposited above the NbSe$_2$ flake to protect it from oxidization. The optical micrograph of the device is shown in Fig.~\ref{fig1}(a). All the measurements are performed in a $^3$He refrigerator at bath temperatures $4.2\,\text{K}$~(gas) and $0.7\,\text{K}$ (liquid). Where introduced, magnetic field is perpendicular to the plane of the flake. The details of our noise measurements can be found in Appendix~A.

Out of four gold contact pads, three provide electrical contact to the flake (labelled with C$1$, C$2$ and C$3$). 
Fig.~\ref{fig1}(b) demonstrates temperature ($T$) dependence of the normalized quasi- four-terminal linear-response resistances which excludes the wiring contribution of the setup: $R_{12}$ measured between pads C$1$ and C$2$ (blue circles), and $R_{3}$ measured between pads C$3$ and both pads C$1$ and C$2$ in parallel~(red crosses). At room temperature these resistances are approximately~$800\,\Omega$ and $6\,\text{k}\Omega$, respectively. Upon cooling down, in the vicinity of $T\approx6\,\text{K}$ where the flake enters the superconducting transition region~\cite{Khestanova2018}, both $R_{12}$ and $R_{3}$ first demonstrate steep decrease. Upon further cooling, however, the behavior of $R_{12}(T)$ and $R_{3}(T)$ is different. While $R_{12}(T)$ monotonically decreases down to the lowest available $T=0.7\,\text{K}$, $R_{3}(T)$ starts increasing at around $T\approx5\,\text{K}$, reaching $6\,\text{k}\Omega$ at~$4.2\,\text{K}$ and $17\,\text{k}\Omega$ at~$0.7\,\text{K}$ (the $T$-dependence below~$4.2\,\text{K}$ is not shown). We emphasize that the observed resistance change due to the decrease of~$T$ is approximately~$100\,\Omega$ and by far exceeds the normal resistance of gold electrodes beneath the flake (approximately~$10\,\Omega$). As we argue below, this behavior reflects the formation of a tunnel-type contact between the flake and the gold pad~C$3$. We will show that it is characterized by the average transmission probability $\langle\tau\rangle\approx0.5$ which is not much smaller than unity. Still, we will refer to this contact as to the TJ in contrast to the contacts at the C1(C2)/NbSe2 interfaces. Magnetic field destroys superconductivity of the flake, see the inset of Fig.~\ref{fig1}(b) for the curves of~$R_{12}(B)$ at both $T=4.2\,\text{K}$ and $0.7\,\text{K}$. Here, the observed steps are likely related to the thickness inhomogeneity of the flake. This inhomogeneity is most clearly manifested in the $T$-dependence of the four-terminal resistance of another thin flake, see SM Fig.~S1(b). At the same time, we do not observe similar step features for the thick flake, see Fig.~S2(b). Additional transport data obtained on another exfoliated thin flake on prepatterned gold contacts as well as data obtained on lithographically patterned thick flake are demonstrated in Supplementary Material Fig.~S1 and Fig.~S2, respectively. We note that generally most of interfaces between prepatterned gold and NbSe$_2$ in our devices turn out to be in the few hundred Ohms range and demonstrate monotonically decreasing $R(T)$ upon cooling down. The fact that some of the contacts demonstrate tunnel characteristics is occasional and is not controlled~\cite{Hoshi2019}. In Supplementary Material Fig.~S3 we show one more example of a tunnel contact on another thin NbSe$_2$ flake.

\section{Tunnel junction characterization}
\begin{figure}[h]
\begin{center}
\includegraphics[width=\columnwidth]{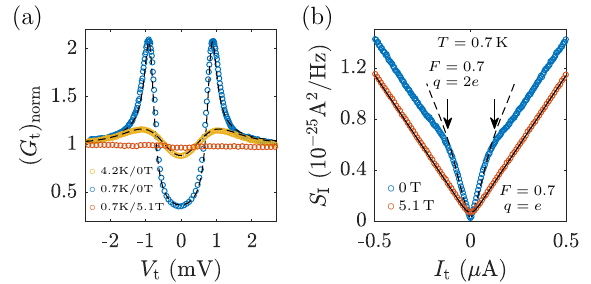}
\end{center}
\caption{(a)~Normalized differential conductance of the tunnel probe. The dashed lines are fits according to the BTK model using $\Delta=0.69\,\text{meV}$, $Z=1.03$ at $4.2\,\text{K}$ and $\Delta=0.88\,\text{meV}$, $Z=1.12$ at $0.7\,\text{K}$. (b)~Current noise spectral density of the tunnel probe at $T=0.7\,\text{K}$ in a magnetic field of $B=0$ and $5.1\,\text{T}$. Dashed lines are the fits with $F=0.7$ and $q=e$ in $B=5.1\,\text{T}$ and $q=2e$ in $B=0\,\text{T}$ (see text).}
\label{fig2}
\end{figure}

Using the schematics from the upper part of Fig.~\ref{fig1}(c), we now characterize transport and noise properties of the~TJ. Here, we can safely neglect the contribution to the measured resistance from both interfaces C$1$(C$2$)/NbSe$_2$ and from the flake itself since these resistances are limited by no more than approximately~$100\,\Omega$, see the inset of Fig.~\ref{fig1}(b). In high enough magnetic fields, TJ's $I$-$V$ curve is linear at both $T=4.2\,\text{K}$ and $0.7\,\text{K}$ so that the differential conductance, $G_{\text{t}}=dI_{\text{t}}/dV_{\text{t}}$, is bias-independent, see red circles in Fig.~\ref{fig2}(a) for the normalized to high bias voltage data at $T=0.7\,\text{K}$ in the magnetic field $B=5.1\,\text{T}$. Upon decreasing~$B$, $G_{\text{t}}(V_{\text{t}})$ starts demonstrating features inherent to NS-contacts in the tunnelling regime. Namely, we observe the peaks in $G_{\text{t}}(V_{\text{t}})$ curves which are most pronounced in $B=0$ (blue circles). Moreover, these peaks smooth out upon increasing temperature (yellow circles). The corresponding $IV$-curves are demonstrated in Supplementary Material Fig.~S4.

Quantitative description of differential conductance curves of NS-contacts is usually performed by comparison to the behavior predicted by the BTK-model~\cite{Blonder1982},
\begin{equation*}
I_{\text{NS}}\propto\int\limits_{-\infty}^{+\infty}\left[f_0(\varepsilon-eV)-f_0(\varepsilon)\right]\left[1+A(\varepsilon)-B(\varepsilon)\right]\,d\varepsilon,
\end{equation*}
where $A(\varepsilon)$ and $B(\varepsilon)$ are the Andreev and normal reflection probabilities listed in Table~II of ref.~\cite{Blonder1982} and depending on the dimensionless number~$Z$ which specifies the strength of the interface potential barrier. We observe that the above expression provides reasonable agreement with the experimental data, see the  dashed lines in~Fig.~\ref{fig2}(a). The used fitting parameters are $\Delta=0.69\,\text{meV}$, $Z=1.03$ at $4.2\,\text{K}$ and $\Delta=0.88\,\text{meV}$, $Z=1.12$ at $0.7\,\text{K}$. We note that such obtained $Z$ is almost $T$-independent, and the values of superconducting gap potential~$\Delta$ at two temperatures fall on the expected $T$-dependence of superconducting gap, $\Delta\propto\tanh\left(1.74\sqrt{T_{\text{c}}/T-1}\right)$, with the superconducting transition temperature~$T_{\text{c}}=5.9\,\text{K}$. Overall, the data of~Fig.~\ref{fig2}(a) indicate that the C$3$/NbSe$_2$ interface is represented by a TJ which is characterized by the set of energy-independent transmission probabilities which on the average may be estimated as~$\langle\tau\rangle\sim(1+Z^2)^{-1}\approx0.5$.

Additional information is provided by the standard noise measurements~\cite{Blanter2000}. Here, we pass the current~$I_{\text{t}}$ across the TJ like in differential conductance measurements, see the upper part of Fig.~\ref{fig1}(c), and measure the voltage fluctuations utilizing the resonant tank circuit technique with the central frequency of approximately~$15\,\text{MHz}$. The results expressed in terms of the current noise spectral density, $S_{\text{I}}$, as a function of~$I_{\text{t}}$, are demonstrated in~Fig.~\ref{fig2}(b). At $T=0.7\,\text{K}$, in high enough magnetic field $B=5.1\,\text{T}$ where the TJ's $I$-$V$ characteristics is linear~(red circles), we observe linear growth of~$S_{\text{I}}$ with the absolute value of~$I_{\text{t}}$ which is the characteristic feature of the shot noise. In the whole range of TJ~currents, the data are perfectly consistent (see the solid black line) with the standard shot noise expression 
\begin{equation}
S_{\text{I}}=4k_{\text{B}}TG+2q|I|F(\coth\xi-1/\xi),
\label{stdexpr}
\end{equation}
where 
$G=G_{\text{t}}$ is the TJ differential conductance, $I=I_{\text{t}}$ is the current through the TJ,
$q=e$ is the electron charge, $\xi=q|V|/(2k_{\text{B}}T)$ is determined by the TJ bias voltage $V=V_{\text{t}}$,  and the Fano-factor $F=0.7$. In the scattering matrix formalism it is given by $F = \sum\tau_i (1-\tau_i)/ \sum\tau_i$, where $\tau_i$ are the transmission probabilities of conduction eigen-channels of a given conductor. Typically, $F$ varies between $0$ and $1$, reaching $F=0$ on the conductance plateaus of quantum point contacts and $F=1$ for TJs with all $\tau_i \ll1$. Note that for the single-channel TJ, based on the estimate from the BTK fits for the differential conductance, one should have expected $F=1-\tau\approx0.5$. Important for further discussion is the linear dependence of $S_{\text{I}}(I_{\text{t}})$ revealing the lack of electron-phonon relaxation~\cite{Nagaev1992} in transport across the TJ. 

In zero magnetic field, the $S_{\text{I}}(I_{\text{t}})$-curve (blue circles) demonstrates two kinks at currents corresponding to the bias voltages $(V_{\text{t}})_{\text{kink}}\approx\pm\Delta/e=0.88\,\text{mV}$, see arrows in~Fig.~\ref{fig2}(b). Quantitatively, the low current data are reasonably described using the same expression~(\ref{stdexpr}) but now with $q=2e$, while the slope of the high current data is the same as for the case of $B=5.1\,\text{T}$. This behavior is expected~\cite{Beenakker1997} and reflects the Andreev reflection mechanism in transport across NS~contacts~\cite{Jehl2000,Kozhevnikov2000,Das2012,Ronen2016,Denisov2022}: at sub-gap voltages the current is converted directly into supercurrent which is accomplished by the reflection of a hole back into the metal so that charge~$2e$ is transferred across the interface. 

Altogether, the results presented in Fig.~\ref{fig2} -- differential conductance like that of a TJ which can be reasonably described using the BTK theory, the effective charge doubling demonstrated in the shot noise in zero magnetic field, and the value of Fano-factor which is in reasonable agreement with the BTK fits -- in physical terms clearly show the formation of a TJ at the C3/NbSe2 interface.
Using the value of the normal state low-temperature  TJ resistance $R_{\text{N}}\approx5\,\text{k}\Omega$, we estimate the number of spin-degenerate conduction channels across the C$3$/NbSe$_2$ interface as $N= h/(2e^2\langle\tau\rangle R_{\text{N}})\approx5$. The obtained result corresponds, most likely, to a point contact with the cross-section area being on the order of~$1\,\text{nm}^2$.

\section{Good contacts characterization}
\begin{figure}[h]
\begin{center}
\includegraphics[width=\columnwidth]{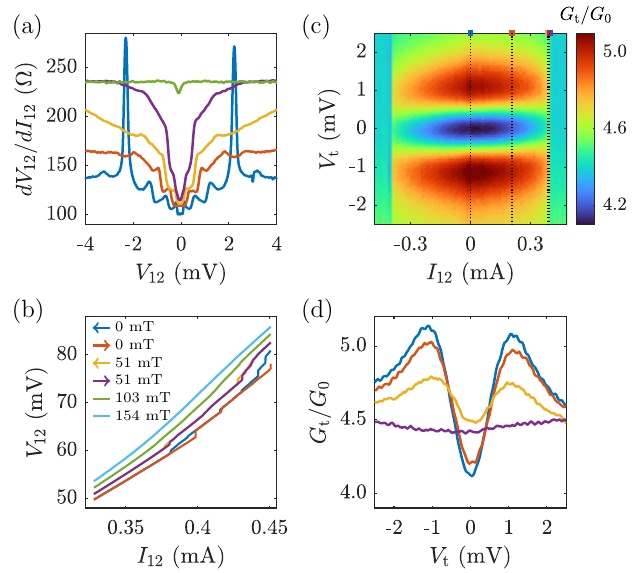}
\end{center}
\caption{(a)~Differential resistance of the flake as a function of bias voltage in magnetic fields of~$B=0,\,2.1,\,3.1,\,4.1$ and $5.7\,\text{T}$ (from blue to green curves). (b)~$I$-$V$ curve of the flake demonstrating the voltage steps and their evolution in a magnetic field at $T=4.2\,\text{K}$. (c)~Differential conductance of the tunnel probe as a function of flake current~$I_{12}$ and the tunnel probe bias voltage at~$T=4.2\,\text{K}$. (d)~Linecuts of differential conductance at specific values of~$I_{\text{flake}}=0,\,0.21,\, 0.389$ and $0.397\,\text{mA}$ indicated by dashed lines on panel~(c).}
\label{fig3}
\end{figure}

Unlike C$3$/NbSe$_2$ interface, both interfaces C$1$(C$2$)/NbSe$_2$ are relatively transparent so we qualify them as the good contacts. In~Fig.~\ref{fig3}(a) we plot $B$-evolution of the differential resistance $dV_{12}/dI_{12}$ at low biases, measured at~$T=0.7\,\text{K}$. The blue curve, corresponding to $B=0$, exhibits a set of resistance peaks. These peaks are absent at~$T=4.2\,\text{K}$ (see Supplementary Material Fig.~S5) and smooth out at increasing~$B$. We note that single peaks at few~mV bias voltages similar to our biggest peaks were also observed in~\cite{Paradiso2019} at intermediate~$T=3.3\,\text{K}$. As we demonstrate below, even at the lowest available $T=0.7\,\text{K}$ electron-phonon scattering length~$l_{\text{e-ph}}$ in NbSe$_2$ is much shorter than the distance between two interfaces so that the device should be considered as a series connection of two independent NS~interfaces and the flake itself. In our device it is not possible to reliably separate contribution of two interfaces to~$R_{12}$. We note that for the single graphene/NbSe$2$ interface, the recent paper~\cite{Moriya2020} identified similar peaks as originating from the transport between the superconductor and the proximitized conductor underneath, in our case corresponding to the transport across the proximitized Au/NbSe$_2$ interface. The occurence of multiple peaks may be related to the varying NbSe$_2$ thickness along the interface. The corresponding evolution of differential resistance~$R_{12}$ with magnetic field and bias voltage is demonstrated in Supplementary Material Fig.~S6.

Similar to the findings of~\cite{Moriya2020}, we also observe the voltage steps in the current-voltage characteristics~$V_{12}(I_{12})$, at large bias currents around~$0.4\,\text{mA}$ (we note that there are no voltage steps at bias currents below $0.35\,\text{mA}$). The corresponding zero-field data are demonstrated in~Fig.~\ref{fig3}(b) by blue and red curves ($B=0$), and yellow and violet curves ($B=51\,\text{mT}$), with two colors in each pair reflecting two opposite current sweep directions. These steps decrease in height and smooth out at relatively small magnetic fields: already at $B=103\,\text{mT}$, the current-voltage characteristics for two opposite sweep directions are not distinguishable. Overall, the observed hysteresis along with the $B$-behavior indicate the superconducting origin of the steps. Qualitatively similar hysteretic behavior was recently reported for the geometrically well-defined system represented by a semiconducting full-shell epitaxial Al-InAs~nanowire connected on its ends to two thick Au~pads~\cite{Shpagina2024}. It was demonstrated that observation of superconducting state at bias voltages~$V\gg\Delta/e$ implies strong energy relaxation. Compared to the one-dimensional wire-shaped geometry of~\cite{Shpagina2024}, in our case poorly defined two-dimensional geometry for current redistribution probably results in multiple hysteretic steps in the $I$-$V$~characteristics rather than to a single hysteretic loop. Yet another important difference is the stronger electron-phonon relaxation in NbSe$_2$ compared to Al (see below), leading to observation of superconducting state even further above the superconducting gap.

\section{Conductance spectroscopy}
Evolution of local superconducting state in the flake in the presence of bias current~$I_{12}$ can be visualized by the widely accepted conductance tunnel probe spectroscopy~\cite{Blonder1982,Anthore2003,LeSueur2008,Vercruyssen2012,Menard2015,Dvir2018,Jalabert2023} using the TJ at the C$3$/NbSe$_2$ interface, see the schematics of the bottom part of~Fig.~\ref{fig1}(c). In this experiment, we measure the differential resistance, $R_{\text{t}}=dV_{\text{t}}/dI_{\text{t}}$, of the TJ. Fig.~\ref{fig3}(c) demonstrates the color-scale plot of the differential conductance, $G_{\text{t}}=1/R_{\text{t}}$, of the tunnel probe as a function of the tunnel probe bias voltage~$V_{\text{t}}$ and of the current through the flake~$I_{12}$, measured at~$T=4.2\,\text{K}$ 
with the sweep of~$I_{12}$ from positive to negative values. Note that the current~$I_{\text{t}}$ flowing into the flake through the TJ doesn't exceed $0.5\,\mu\text{A}$ and is negligible on the scale of~$I_{12}$. The shaded squares with the corresponding dashed black lines in Fig.~\ref{fig3}(c) indicate the specific values of~$I_{12}$ for which $G_{\text{t}}(V_{\text{t}})$ are shown in Fig.~\ref{fig3}(d). Below the transition value of~$I_{12}\approx0.39\,\text{mA}$, at decreasing the magnitude of $I_{12}$, the tunnel probe differential conductance gradually evolves in a way so that the conductance peaks move to larger values of~$|V_{\text{t}}|$ and the zero-bias conductance dip becomes more pronounced. This indicates the continuous decrease of the local electronic temperature~$T_{\text{e}}$ in the vicinity of the~TJ. The result of Fig.~\ref{fig3}(c) allows us to demonstrate the qualitative difference of our case compared to the situation where one can neglect electronic energy relaxation. Namely, we note that the transition between local superconducting and normal states happens in a similar way regardless of whether $|I_{12}|$ decreases or increases.
In contrast, for the Al~wire shorter than electron-phonon scattering length and connected between two large copper pads~\cite{Vercruyssen2012}, evolution of superconducting state was shown to differ depending on the current sweep direction. The difference comes from the realization in the case of negligible energy relaxation~\cite{Vercruyssen2012} of the so-called global and bimodal superconducting states for two opposite current sweep directions. The global state is one coherent superconducting state extending over the full length of the wire and realized for the increasing current sweep, while the bimodal state consists of two separate superconducting domains located at each cold end of the wire and realized for the decreasing current sweep. Strong energy relaxation in NbSe$_2$ makes this picture inapplicable in our case. We were not able to perform similar measurements at $0.7$\,K since at low~$T$ the device was generating some noise overloading the preamplifier at large values of~$I_{12}$.

As a final remark, note that in the above discussion we assumed the current density to be much smaller than the depairing current density of NbSe$_2$ so that the evolution of $G_{\text{t}}(I_{12})$ was attributed exclusively to the changes in local electronic temperature~$T_{\text{e}}$. Unfortunately, the poor defined geometry of current distribution in our device doesn’t allow to judge on the possible influence of bias current~$I_{12}$ on the density of states in NbSe$_2$~\cite{Anthore2003}.
\section{delta-$T$ noise measurements}
One way to experimentally quantify electron-phonon relaxation in a device is to measure its current noise~\cite{Roukes1985,Steinbach1996,Huard2007,Betz2013}. Namely, for the two-terminal diffusive conductor without electron-phonon scattering its shot noise is described by~(\ref{stdexpr}) with $F=1/3$ and $F=\sqrt3/4$ for the cases of negligible and strong electron-electron scattering, respectively. For the diffusive conductor, suppression of the current noise below $F=1/3$ indicates the relevance of electron-phonon scattering which is characterized by electron-phonon coupling~$\Sigma_{\text{e-ph}}$ entering to the power flow from electron to phonon subsystem via
\begin{equation}
P_{\text{e-ph}}={\mathcal V}\Sigma_{\text{e-ph}}\left(T_{\text{e}}^n-T_{\text{ph}}^n\right),
\label{eph}
\end{equation}
where ${\mathcal V}$ is the system volume, $T_{\text{e}}$ and $T_{\text{ph}}$ are electronic and phononic temperatures, and exponent~$n$ typically varies in the range $n\approx3-5$~\cite{Giazotto2006}. In the steady state $P_{\text{e-ph}}$ is exactly the released Joule heat power. The relation~(\ref{eph}) works best for the case of long enough devices where $T_{\text{e}}$ and $T_{\text{ph}}$ can be considered as position-independent everywhere besides short regions near the terminals. From the experimental point of view, special care should be paid to ensure that contact pads and the corresponding interfaces do not contribute to the measured noise signal~\cite{Polyak2024d}. If the device under study is a superconductor connected to normal contacts, taking contacts into account turns out to be a fundamental difficulty which can not be overcome in the standard two-terminal noise thermometry. In the present case of a thin NbSe$_2$ flake which degrades during the standard lithography techniques, the proper quality of contacts can hardly be ensured for the thermometry experiment even for the flake in the normal state.

Below we will perform a local delta-$T$ noise experiment which is advantageous~\cite{Tikhonov2016,Lumbroso2018,Tikhonov2020} in this situation, see the schematics from the bottom part of Fig.~\ref{fig1}(c). Consider a~TJ characterized by energy-independent transmission probabilities~$\tau_n$ and differential conductance~$G_{\text{t}}$, which connects two conductors: the one in equilibrium with Fermi-Dirac energy distribution~(ED), $f_0(\varepsilon,T_{0})=[\exp(\varepsilon/k_{\text{B}}T_{0})+1]^{-1}$, and the nonequilibrium one with some~ED $f_{\text{NEQ}}(\varepsilon)$ which is generally position-dependent. Here $T_0$ is the bath temperature and the two conductors in our case are represented by the gold pad~C$3$ and the NbSe$_2$ flake, respectively, see the lower part of Fig.~\ref{fig1}(c) with the shaded region representing the~TJ. Generally, the noise temperature of a TJ under applied bias voltage~$V_{\text{t}}$, $T_{\text{N}}(V_{\text{t}})=S_{\text{I}}/4k_{\text{B}}G_{\text{t}}$, allows one to extract~\cite{Gramespacher1999,Ota2017,Tikhonov2020}
\begin{equation}
f_{\text{NEQ}}(\varepsilon)=\frac12-\frac1F\frac{d\left(k_{\text{B}}T_{\text{N}}\right)}{d(eV_{\text{t}})}\Big|_{eV_{\text{t}}=\varepsilon}.
\label{spectroscopy}
\end{equation}
Here $S_{\text{I}}(V_{\text{t}})$ is the TJ's~current noise spectral density, $F=\sum \tau_n(1-\tau_n)/\sum\tau_n$ is the Fano-factor of the TJ and $V_{\text{t}}$ is the bias voltage on it. This relation is valid provided $T_0$ is much smaller than the characteristic energy scale on which~$f_{\text{NEQ}}(\varepsilon)$ changes significantly. 

The important special case is realized for nonequilibrium conductors with strong electron-electron or electron-phonon relaxation so that $f_{\text{NEQ}}(\varepsilon)$ is of Fermi-Dirac type, $f_{\text{NEQ}}(\varepsilon)=f_0(\varepsilon,T_{\text{e}})\equiv[\exp(\varepsilon/k_{\text{B}}T_{\text{e}})+1]^{-1}$, with nonequilibrium electronic temperature~$T_{\text{e}}$. In this particular case, the local value of $T_{\text{e}}$ can be extracted from the noise measurements using~\cite{Blanter2000}
\begin{equation}
T_{\text{N}}=\frac{T_{\text{e}}+T_0}{2}+\frac{F}{2k_{\text{B}}}\int d\varepsilon \left[f_0(\varepsilon,T_{\text{e}})-f_0(\varepsilon,T_0)\right]^2,
\label{getTloc}
\end{equation}
applicable, in contrast to~(\ref{spectroscopy}), at any value of~$T_{\text{e}}$. Note there is no average current through the~TJ and the nonequilibrium noise comes from different temperatures on two sides of the~TJ. Below we will use~(\ref{getTloc}) to obtain the local value of~$T_{\text{e}}$ in the vicinity of the~TJ in the driven out-of-equilibrium NbSe$_2$ flake from the measurements of~$T_{\text{N}}$. Note that this relation assumes the condition $G_{\text{t}}\ll R_{12}^{-1}$ which assures the negligible heat conduction across the TJ.

\begin{figure}[!h]
\begin{center}
\includegraphics[width=\columnwidth]{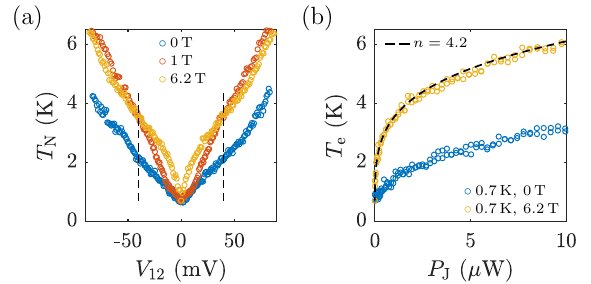}
\end{center}
\caption{(a)~Noise temperature of the~TJ as a function of bias voltage on the flake in $B=0,\,1$ and $6.2\,\text{T}$ measured at bath temperature~$T_0=0.7\,\text{K}$. Dashed lines indicate the range of bias voltages used for the analysis in panel~(b). (b)~Local electronic temperature in NbSe$_2$ under the TJ as a function of the total Joule power released in the flake due to the current~$I_{12}$. Dashed line is a fit using~(\ref{eph}) with $n=4.2$ and ${\mathcal V}\Sigma_{\text{e-ph}}=5\cdot10^{-9}\,\text{W}/\text{K}^{4.2}$.}
\label{fig4}
\end{figure}

In the experiment we measure voltage fluctuations of a TJ in response to the bias current through the flake. We concentrate on the low-$T$ data obtained at $T=0.7\,\text{K}$ where the exponent~$n$ entering~(\ref{eph}) can be extracted unambiguously due to the relatively large range of accessible~$T_{\text{N}}$. The data expressed in terms of noise temperature~$T_{\text{N}}$ of a~TJ as a function of bias voltage~$V_{12}$ on the flake are demonstrated in Fig.~\ref{fig4}(a) for the three representative values of magnetic field.  The main effect of~$B$ on the $T_{\text{N}}(V_{12})$ curves occurs in the bias range below approximately~$40\,\text{mV}$ (see the vertical dashed lines) where the slope of the curves grows significantly with the increasing magnetic field. Further, we limit ourselves to this region of $|V_{12}|\lesssim 40\,\text{mV}$ where the dependencies $T_{\text{N}}(V_{12})$ are sublinear. As is evident from Fig.~\ref{fig4}(a), at larger bias voltages these dependencies tend to linear ones which may indicate mediated by amorphous SiO$_2$ relaxation bottleneck with~$n\approx2$~\cite{Baeva2020,Baeva2021}. 

Using~(\ref{getTloc}), from the data of Fig.~\ref{fig4}(a) we extract the local electronic temperature in the NbSe$_2$ flake in the vicinity of the~TJ. The results are demonstrated in~Fig.~\ref{fig4}(b). In $B=6.2\,\text{T}$, the flake is in the normal state, see the inset of Fig.~\ref{fig1}(b), and the power flow between electron and phonon subsystems is well described by~(\ref{eph}) with $T_0=0.7\,\text{K}$, $n=4.2$ and ${\mathcal V}\Sigma_{\text{e-ph}}=5\cdot10^{-9}\,\text{W}/\text{K}^{4.2}$, see yellow symbols and the corresponding black dashed fit in~Fig.~\ref{fig4}(b). This value allows one to estimate the electron-phonon scattering length in NbSe$_2$ via~\cite{Denisov2020}
\begin{equation*}
l_{\text{e-ph}}=L\left[{\cal L}/nT^{n-2}{\mathcal V}\Sigma_{\text{e-ph}}R_{12}\right]^{1/2},
\end{equation*}
where $L$ is the effective device length and ${\cal L}=2.44\cdot10^{-8}\,\text{W}\Omega\text{K}^{-2}$ is the Lorenz number. Importantly, even at the lowest available $T_0=0.7\,\text{K}$, $l_{\text{e-ph}}$ is approximately 10~times shorter than the effective device length, $l_{\text{e-ph}}/L\approx0.1$, demonstrating the importance of energy relaxation in any nonequilibrium experiment with NbSe$_2$ on a micrometer scale. Quantitatively, the effective device length in our case can be roughly estimated on the order of~$10\,\mu\text{m}$ implying 
$l_{\text{e-ph}}\approx0.8/T^{1.1}\,\mu\text{m}\,\text{K}^{1.1}$.
It justifies the assumption made about the Fermi-Dirac shape of~ED in the flake region near the C3/NbSe$_2$ interface. Note that observation of $n\approx4$ excludes diffusion of heat into metallic contact pads as a dominant relaxation mechanism in the considered energy range which would be manifested in the observation of $n=2$~\cite{Henny1999}. We also note that observation of $n\approx4$ may reflect the 2D~nature of acoustic phonons similar to the case of graphene~\cite{Kubakaddi2009,Baker2012,Betz2012,Fong2012}. The fact that the experimental data are best fitted with $n$ slightly different from an integer value may be due to the fact that the model of 2D phonons is not completely applicable.

The above analysis implies that Joule heat is predominantly released in the NbSe$_2$ flake itself. However, from the magnetoresistance data presented in the inset of Fig.~\ref{fig1}(c) one can etimate that in the normal state the flake resistance is only $1.5$~times greater than the total interfaces resistance. We therefore deduce that our result allows for the spread $\mathcal{V}\Sigma_{\text{e-ph}}\approx(3$$-$$5)\cdot10^{-9}\,\text{W}/\text{K}^{4.2}$. From here, $l_{\text{e-ph}}$~may be  approximately $30\%$ greater than estimated above. Still, we emphasize that our result for electron-phonon relaxation length in NbSe$_2$ in the normal state, $l_{\text{e-ph}}(\text{NbSe}_2,\,0.7\,\text{K})\approx1\,\mu\text{m}$, is significantly smaller than the recent results~\cite{Pinsolle2016} obtained for the normal state Al at the same~$T$. Namely, acquired by the measurements of the dynamical response of thermal noise to an ac excitation, the values of~$D$, $A$ and $n$ presented in Table~I of ref.~\cite{Pinsolle2016} suggest that $l_{\text{e-ph}}(\text{Al},\,0.7\,\text{K})$ lies in the range $3\,\mu\text{m}$$-$$9\,\mu\text{m}$ for the relatively low-quality~Al with the diffusion coefficient $D\lesssim50\,\text{cm}^2/\text{s}$. Even bigger values are expected for gold where $l_{\text{e-ph}}(\text{Au},\,0.7\,\text{K})\approx14\,\mu\text{m}$~\cite{Baeva2021}. Our results for the normal state NbSe$_2$ also suggest the weaker $T$-dependence of the electron-phonon scattering rate, $\tau_{\text{e-ph}}(\text{NbSe}_2)\propto T^{-2.2}$, compared to the clean metallic three-dimensional case where one would expect the $\propto T^{-3}$-dependence.

In zero magnetic field, the NbSe$_2$ flake is superconducting so that Joule heat due to current~$I_{12}$ releases exclusively at C$1$(C$2$)/NbSe$_2$ interfaces. In this case, since the distance between interfaces and the tunnel probe is few times greater than electron-phonon scattering length determined in the normal state, the increase of local electronic temperature nearby the probe at a given value of $P_{\text{J}}$ may be expected to be smaller compared to the normal state case. This is indeed what we observe, see blue symbols in Fig.~\ref{fig4}(b). In contrast to the normal state case, temperature increase is now detected not where heat is released so that the role of heat conduction increases. Whether heat conduction in our device occurs by quasiparticle transport in NbSe$_2$ or by phonons in either of the 2D~crystals NbSe$_2$ and~hBN is not possible to conclude from our experiment and requires better defined geometry. Finally, we note that the increase of energy relaxation length in superconductors at low temperatures compared to the normal case~\cite{Reizer}, is expected to be negligible in our device at $T\gtrsim1\,\text{K}$.

\section{Conclusions}
In conclusion, we performed transport and noise measurements for a thin superconducting NbSe$_2$ flake laid onto the predefined gold contacts and covered with a thin hBN~flake. Tunnel junction formed at one of the NbSe$_2$/Au interfaces was utilized in two ways. First, by observation of effective charge doubling in the shot noise of this TJ at sub-gap energies, we identified Andreev reflection regime corresponding to tunneling across NS~contact. Second, releasing Joule heat power on the NbSe$_2$ side of the TJ and measuring the resulting delta-$T$ noise of the~TJ, we extracted electron-phonon scattering length in~NbSe$_2$ to be
$l_{\text{e-ph}}\approx0.8/T^{1.1}\,\mu\text{m}\,\text{K}^{1.1}$. 
 Given the introduced approach is applied to geometrically well-defined superconducting devices, it is promising in the study of nonequilibrium configurations.

\section*{Acknowledgments}
We thank D.V.~Shovkun for helpful discussions.

\section*{Funding}
This work was partially financed under the state task of Osipyan Institute of Solid State Physics Russian Academy of Sciences.

\section*{Conflict of interest}
The authors of this work declare that they have no conflicts of interest.

\appendix
\section{Noise measurements details}
\begin{figure}[!h]
\begin{center}
\includegraphics[width=\columnwidth]{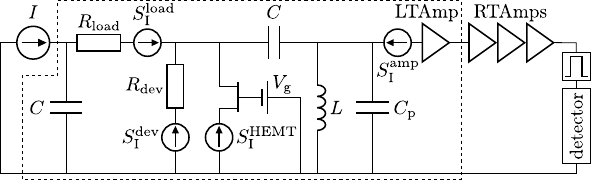}
\end{center}
\caption{Principal noise measurements schematics.}
\label{fig_scheme}
\end{figure}

The principal noise measurements schematics is demonstrated in Fig.~\ref{fig_scheme}. The device ($R_{\text{dev}}$), two load resistors ($R_{\text{load}}=22\,\text{k}\parallel 22\,\text{k}$), the tank circuit with $L=7\,\mu\text{H}$ and $C_{\text{p}}=16\,\text{pF}$ (coaxial cable parasitic capacitance), and the ATF-35143 HEMT are connected in parallel to the input of the high-impedance low-temperature amplifier (LTAmp). The capacitance $C=10\,\text{nF}$ blocks DC current and behaves as a short at the resonant frequency of $15\,\text{MHz}$. These frequencies are high enough so that the contribution of $1/f$~noise can be safely neglected. The signal is further amplified at room temperature (RTAmps) and the corresponding power is detected in a certain bandwidth near the resonant frequency. Noise signal is measured above the contribution of amplifiers which is stable on the time scale of several hours. This is enough to perform averaging of traces for the data in Fig.~2(b) and Fig.~4 -- each single trace takes approximately $30\,\text{s}$ to be measured and the demonstrated data was averaged over~$\sim100$ traces. Voltage fluctuations at the input of the LTAmp are calculated using the standard formalism~\cite{Blanter2000} and are given by
\begin{equation*}
S_{\text{V}}=\left[S_{\text{I}}^{\text{dev}}+S_{\text{I}}^{\text{load}}+S_{\text{I}}^{\text{HEMT}}+S_{\text{I}}^{\text{amp}}\right]\left|Z_{\text{tot}}\right|^2,
\end{equation*}
so that after amplification, up to the background noise of the amplifiers, the detected power can be written as
\begin{equation*}
P=G(R_{\parallel})\left[S_{\text{I}}^{\text{dev}}+S_{\text{I}}^{\text{load}}+S_{\text{I}}^{\text{HEMT}}+S_{\text{I}}^{\text{amp}}\right].
\end{equation*}
Here, $G(R_{\parallel})$ is the impedance-dependent gain for the current noise which  takes into account all the reactive elements as well as the filter frequency response. While all the reactive elements are fixed, the resistance $R_{\parallel}=(R_{\text{dev}}^{-1}+R_{\text{load}}^{-1}+R_{\text{HEMT}}^{-1})^{-1}$ may depend on the transport current. The measurements are carried out in two stages.
\begin{enumerate}
\item During shot noise and delta-$T$ noise measurements, the HEMT is completely depleted (by applying negative enough~$V_{\text{g}}$ to the gate electrode) and doesn't contribute to the signal:  
\begin{equation}
P(I)=G(R_{\parallel})\left[S_{\text{I}}^{\text{dev}}(I)+4k_{\text{B}}T/R_{\text{load}}+S_{\text{I}}^{\text{amp}}\right],
\label{getnoise}
\end{equation}
where $R_{\parallel}(I)=(R_{\text{dev}}^{-1}(I)+R_{\text{load}}^{-1})^{-1}.$
The current noise of the macroscopic resistors~$S_{\text{I}}^{\text{load}}$ does not depend on~$I$ for the used values of~$I$ and is given by the standard Johnson–Nyquist relation. Provided one knows $G(R_{\parallel})$ and $S_{\text{I}}^{\text{amp}}$, expression~(\ref{getnoise}) allows one to extract the device current noise.
\item We now describe the procedure of calibration which is used to get $G(R_{\parallel})$ and $S_{\text{I}}^{\text{amp}}$. Here, no transport current is present but the HEMT gate voltage is varied
\begin{equation*}
P(I=0)=G(R_{\parallel})\left[4k_{\text{B}}T/R_{\parallel}+S_{\text{I}}^{\text{amp}}\right],
\end{equation*}
where $R_{\parallel}=(R_{\text{dev}}^{-1}+R_{\text{load}}^{-1}+R_{\text{HEMT}}^{-1})^{-1}$. By performing measurements at two temperatures we are able to extract both $G(R_{\parallel})$ and $S_{\text{I}}^{\text{amp}}$. We note that the LTAmp is located $\approx20\,\text{cm}$ above the device and its input current noise~$S_{\text{I}}^{\text{amp}}$ is defined by the operating point rather than by the bath~$T$.
\end{enumerate}

\clearpage
\clearpage
\begin{widetext}
\begin{center}
\textbf{\large Supplementary Material to the article ``Local thermometry of NbSe$_2$ flake with delta-$T$ noise measurements''}
\end{center}
\setcounter{equation}{0}
\setcounter{figure}{0}
\setcounter{table}{0}
\setcounter{page}{1}
\setcounter{section}{0}
\makeatletter
\renewcommand{\theequation}{S\arabic{equation}}
\renewcommand{\figurename}{Fig.~S}%
\makeatletter
\def\fnum@figure{\figurename\thefigure}
\makeatother
\renewcommand{\bibnumfmt}[1]{[S#1]}
\renewcommand{\citenumfont}[1]{S#1}

\section{Another thin flake}
\begin{figure}[ht]
	\includegraphics[width = 0.9\columnwidth]{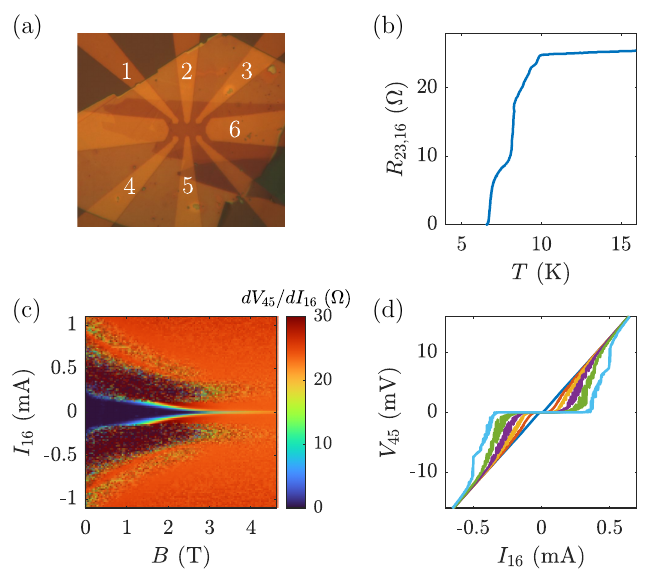}
	\caption{(a)~Optical micrograph of another thin device. Similarly to the main text, exfoliated NbSe$_2$ flake is put on pre-patterned gold contacts and is then covered with hBN flake. (b)~Temperature dependence of the four-terminal linear-response resistance~$R_{23,16}=dV_{23}/dI_{16}$. (c)~Color-scale plot of the four-terminal differential resistance $dV_{45}/dI_{16}$ as a function of the magnetic field and the bias current. (d)~Current-voltage characteristics of data from panel~(c) at specific values of~$B=0,\,0.46,\, 0.97,\, 1.43,\, 1.94$ and $2.91\,\text{T}$.}
	\label{figS1}
\end{figure}
\clearpage
\section{Thick flake}
\begin{figure}[ht]
	\includegraphics[width = 0.9\columnwidth]{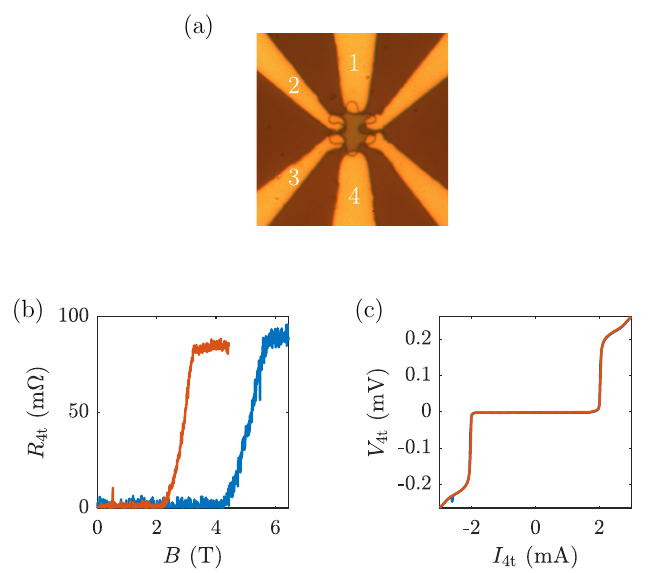}
	\caption{(a)~Optical micrograph of a relatively thick device with lithographically realized contacts. (b)~Magnetic field dependence of a four-terminal resistance at $T=4.2\,\text{K}$ (red curve) and $T=0.5\,\text{K}$ (blue curve). (c)~Four-terminal current-voltage characteristics measured at $T=4.2\,\text{K}$ for two opposite current sweep directions.}
	\label{figS2}
\end{figure}
\clearpage
\section{Another tunnel junction on a thin flake}
\begin{figure}[ht]
	\includegraphics[width = 0.9\columnwidth]{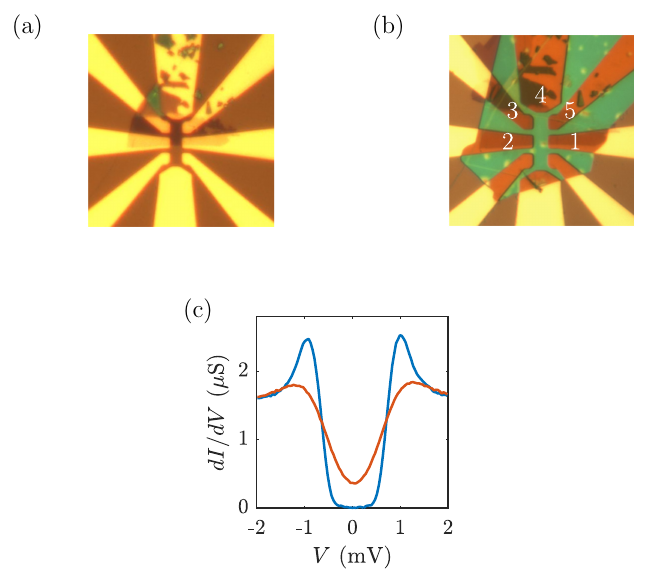}
	\caption{(a,b)~Optical micrograph of a thin device put on predefined gold contacts (a)~before and (b)~after coverage with hBN~flake. Contacts~$1$, $2$, $3$ and $4$ are in the few hundred Ohms range, contact~$5$ demonstrates tunnelling characteristics. (c)~Differential conductance of contact~$5$ at $T=3\,\text{K}$ (red) and $T=0.7\,\text{K}$ (blue).} 
	\label{figTJ}
\end{figure}
\clearpage
\section{$IV$ curves}
\begin{figure}[ht]
	\includegraphics[width = \columnwidth]{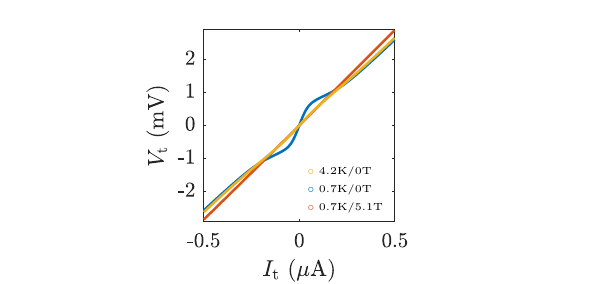}
	\caption{$IV$-curves of the tunnel junction corresponding to the data of Fig.~2(a) of the main text.}
	\label{figiv}
\end{figure}
\clearpage
\section{Differential resistance}
\begin{figure}[ht]
	\includegraphics[width = 0.5\columnwidth]{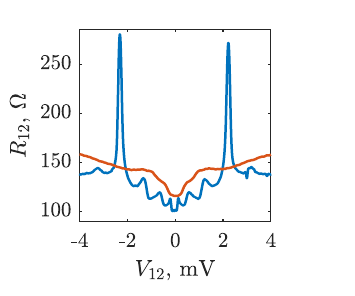}
	\caption{Quasi- four-terminal differential resistance of the device from the main text at~$T=4.2\,\text{K}$ (red curve) and at~$T=0.7\,\text{K}$ (blue curve). The peaks are completely absent at~$T=4.2\,\text{K}$.}
	\label{figS3}
\end{figure}
\begin{figure}[ht]
	\includegraphics[width = 0.5\columnwidth]{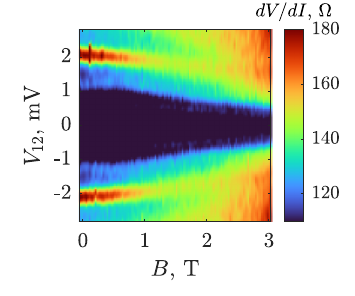}
	\caption{Evolution of differential resistance~$R_{12}$ of the device from the main text with magnetic field and bias voltage at~$T=0.7\,\text{K}$.}
	\label{figS4}
\end{figure}

\end{widetext}
\end{document}